# A Novel Approach of Color Image Hiding using RGB Color planes and DWT

Nilanjan Dey
Asst. Professor Dept. of IT,
JIS College of Engineering
Kalyani, West Bengal, India.

Anamitra Bardhan Roy
B.Tech Student, Dept. of CSE
JIS College of Engineering
Kalyani, West Bengal, India.

Sayantan Dey
B.E. Student, Dept. of IT,
University Institute of Technology
Burdwan, West Bengal, India.

## ABSTRACT

This work proposes a wavelet based Steganographic technique for the color image. The true color cover image and the true color secret image both are decomposed into three separate color planes namely R, G and B. Each plane of the images is decomposed into four sub bands using DWT. Each color plane of the secret image is hidden by alpha blending technique in the corresponding sub bands of the respective color planes of the original image. During embedding, secret image is dispersed within the original image depending upon the alpha value. Extraction of the secret image varies according to the alpha value. In this approach the stego image generated is of acceptable level of imperceptibility and distortion compared to the cover image and the overall security is high.

## General Terms

Color Image Hiding.

## 1. INTRODUCTION

Steganography [1, 2, 3] is the process of hiding of a secret message within an ordinary message and extracting it at its destination. Anyone else viewing the message will fail to know that it contains secret/encrypted data. The word comes from the Greek word "steganos" meaning "covered" and "graphei" meaning "writing".

LSB [4] insertion is a very simple and common approach to embedding information in an image in special domain. The limitation of this approach is vulnerable to every slight image manipulation.

Converting image from one format to another format and back could destroy information secret in LSBs. Stego-images can be easily detected by statistical analysis like histogram analysis. This technique involves replacing N (least significant bit of each pixel of a container image) with the data of a secret message. Stego-image gets destroyed as N increases. In frequency domain, data can be hidden by using Discrete Cosine Transformation (DCT) [5, 8]. Main limitation of this approach is blocking artifact. Grouping the pixel into 8x8 blocks and transforming the pixel blocks into 64 DCT co-efficient each. A modification of a single DCT co-efficient will affect all 64 image pixels in that block. One of the modern techniques of Steganography is Discrete Wavelet Transformation (DWT) approach [6, 7]. In the DWT spread spectrum based approach, binary secret images are dispersed within selective sub-bands using a pseudo random sequence and a session based key [10]. DWT Huffman Encoding based steganographic approach is a very effective technique for gray image hiding [11]. But the limitation of these techniques is the size of the secret image.

We tried to propose a new methodology to hide a color image within another color cover image using alpha-blending technique for the purpose of security. In this approach the imperceptibility and distortion of the Stego image is acceptable and it is resistant to several attacks.

## 2. DISCRETE WAVELET TRANSFORMATION

The wavelet transformation describes a multi-resolution decomposition process in terms of expansion of an Image onto a set of wavelet basis function. The wavelet transform describes a multi-resolution decomposition process in terms of expansion of an Image onto a set of wavelet basis functions. Discrete Wavelet Transformation having its own excellent space frequency localization properly. Applying DWT in 2D images corresponds to 2D filter image processing in each dimension. The input image is divided into 4 non-overlapping multi-resolution sub-bands by the filters, namely (LL1), (LH1), (HL1) and (HH1). The sub-band (LL1) is processed further to obtain the next coarser scale of wavelet coefficients, until some final scale "N" is reached. When "N" is reached, we'll have 3N+1 sub-bands consisting of the multi-resolution sub-bands (LLN) and (LHX), (HLX) and (HHX) where "X" ranges from 1 until "N". Generally most of the Image energy is stored in these sub-bands.

| $LL_3$ | $HL_3$ | $HL_2$ | $HL_1$ |
| $LH_3$ | $HH_3$ | | |
| $LH_2$ | $HH_2$ | | |
| $LH_1$ | | $HH_1$ | |

**Figure 1. Three phase decomposition using DWT.**

The Forward Discrete Wavelet Transform is very suitable to identify the areas in the cover image where a secret image can be embedded effectively due to its excellent space-frequency localization properties. In particular, this property allows the exploitation of the masking effect of the human visual system such that if a DWT co-efficient is modified, it modifies only the region corresponding to that coefficient. The embedded secret image in the lower frequency sub-bands (LLX) may degrade the image significantly, as generally most of the Image energy is stored in these sub-bands. Embedding in the low-frequency sub-bands, however, could increase robustness significantly. In contrast, the edges and textures of the image and the human eye are not generally sensitive to changes in the high frequency sub-bands (HHX). This allows the stego-image to be embedded without being perceived by the human eye. The compromise adopted by many DWT based algorithms, to achieve acceptable performance of imperceptibility and robustness, is to embed the secret image in the middle frequency sub-bands (LHX) or (HLX) and (HHX). The Haar wavelet is also the simplest possible wavelet. Haar wavelet is not continuous, and therefore not differentiable. This property can, however, be an advantage for the analysis of signals with sudden transitions.

## 3. ALPHA BLENDING TECHNIQUE
The way of mixing the two images together to form a final image. Alpha Blending [9] can be accomplished in computer graphics by blending each pixel from the first source image with the corresponding pixel in the second source image.

The equation for executing alpha blending is as follows,

Final pixel = alpha * (First image's source pixel) +
              (1.0-alpha) * (Second image's source pixel)

The blending factor or percentage of colors from the first source image used in the blended image is called the "alpha." The alpha used in algebra is in the range 0.0 to 1.0, instead of 0 to 100%.

Alpha-blending blind Image hiding technique to generate Stego image is given by

Stego Image Embedding:

SII=alpha*(CI) + (1.0-alpha)*(SI)        (1)

Stego Image Extraction:

RSI= (SII - alpha*CI)        (2)

Where, RSI=Recovered Stego Image, SII=Stego image, CI= selected sub-band of the cover image, SI= selected corresponding sub-band of the secret image.

## 4. PROPOSED ALGORITHM
*Secret Image Hiding:*

1. Cover image and Secret image of same size are separated into RGB individual plane

2. RGB individual planes are decomposed into four sub bands (LL, LH, HL and HH) using DWT.

3. RGB individual planes of the secret image are hidden within RGB individual planes of the cover image using alpha blending embedding technique and inverse DWT is applied individually.

4. All three alpha-blended inversed wavelet transformed planes are combined to generate the stego image.

*Secret Image Extraction:*

1. Cover image and Stego image are separated into RGB individual planes.

2. RGB individual planes are decomposed into four sub bands (LL, LH, HL and HH) using DWT.

3. The RGB individual planes of the secret image is extracted from the RGB individual planes of the stego image using alpha blending extraction technique and inverse DWT is applied individually.

4. All three alpha-blended inversed wavelet transformed planes are combined to generate the final extracted true colour secret image.

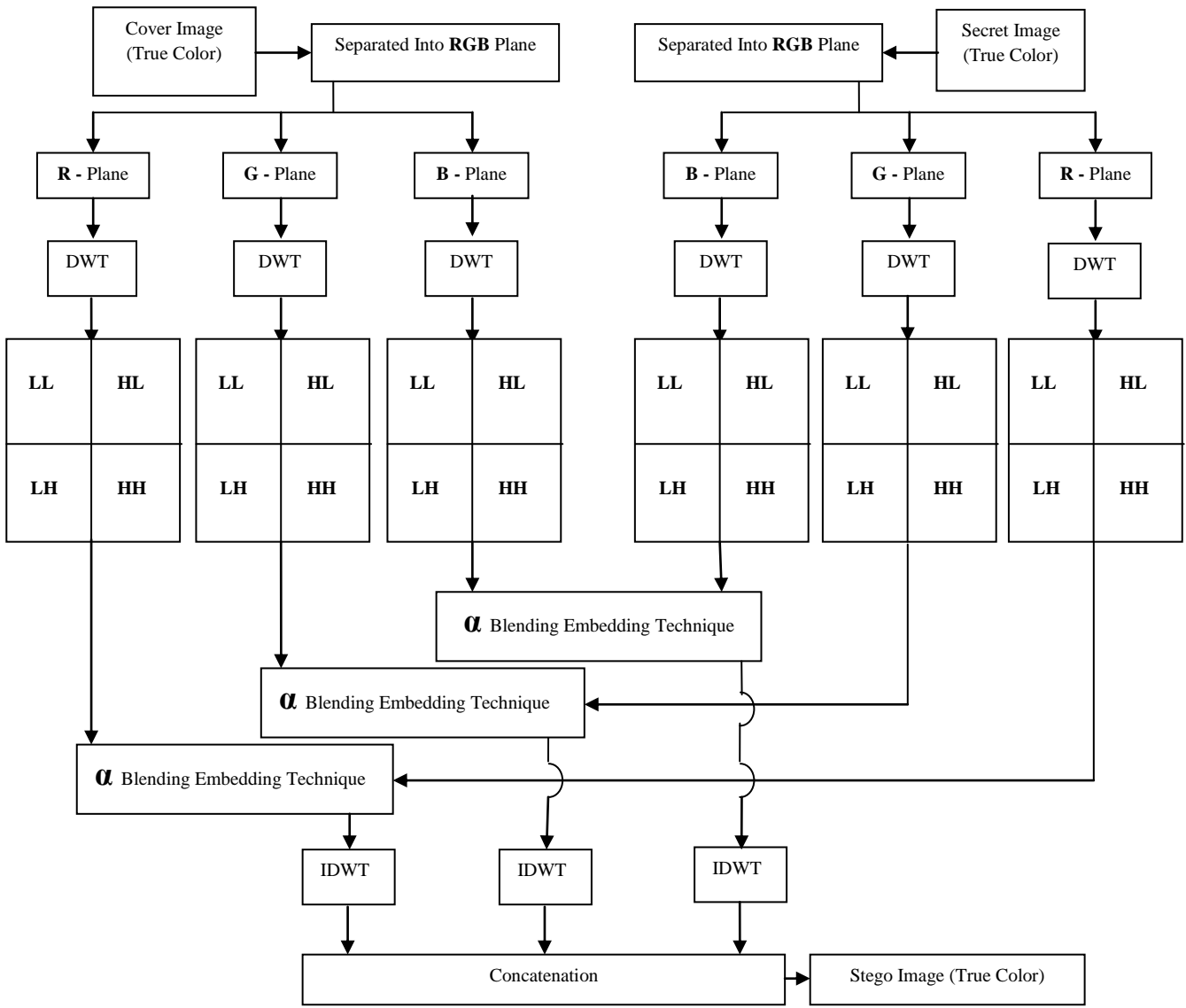

**Figure 2: Secret Image Hiding Technique**

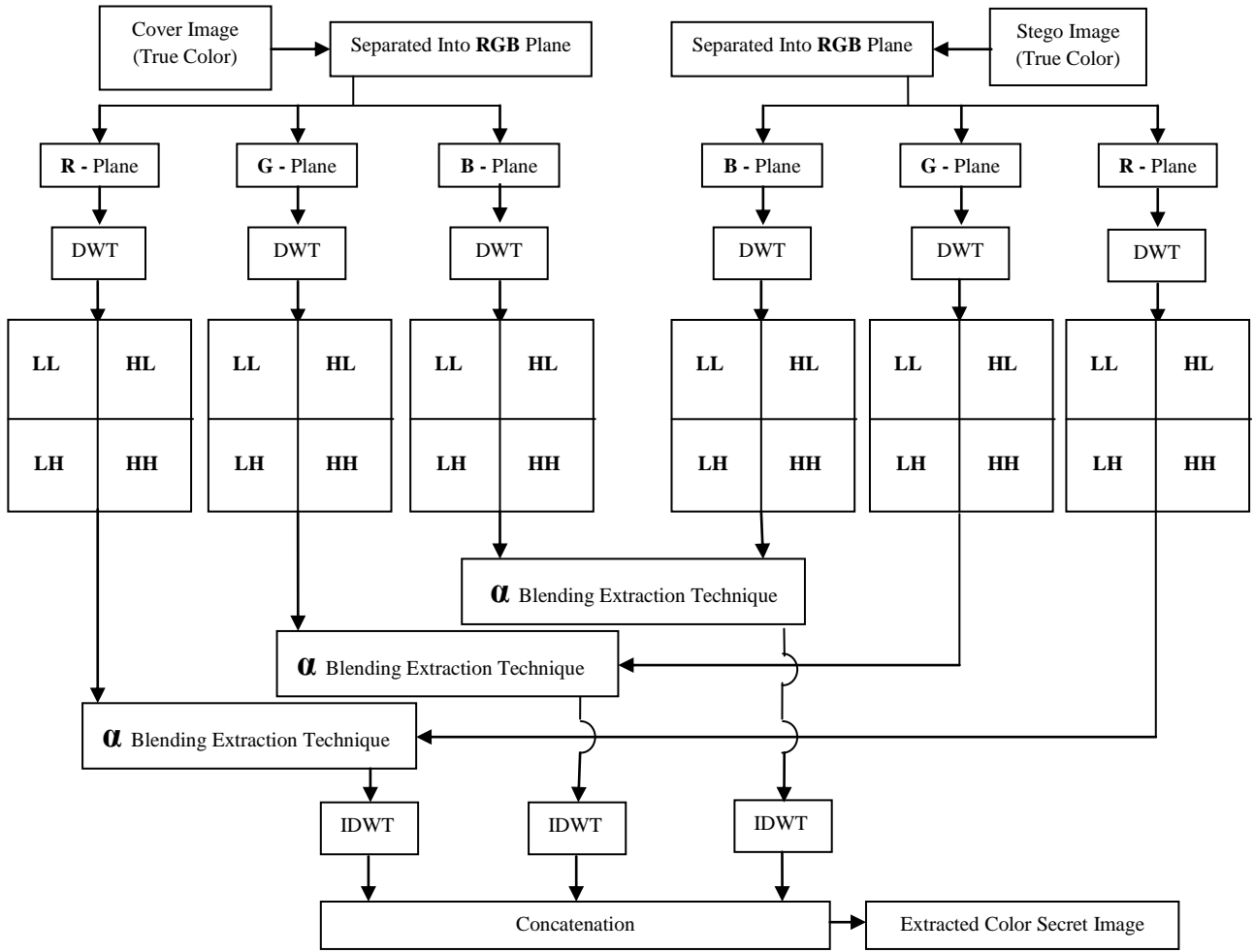

**Figure 3: Secret Image Extraction Technique**

## 5. RESULT AND DISCUSSIONS

Signal-to-noise ratio can be defined in a different manner in image processing where the numerator is the square of the peak value of the signal and the denominator equals the noise variance. Two of the error metrics used to compare the various image de-noising techniques is the Mean Square Error (MSE) and the Peak Signal to Noise Ratio (PSNR).

**Mean Square Error (MSE):**

Mean Square Error is the measurement of average of the square of errors and is the cumulative squared error between the noisy and the original image. MSE is given by the following equation.

$$MSE = \frac{1}{N} \sum_{j,k} (f[j,k] - g[j,k])^2 \quad \ldots \ldots \ldots \ldots \ldots \ldots \ldots (1)$$

**Peak Signal to Noise Ratio (PSNR):**

PSNR is a measure of the peak error. Peak Signal to Noise Ratio is the ratio of the square of the peak value the signal could have to the noise variance.

$$PSNR = 10 \log_{10} \left( \frac{(255)^2}{MSE} \right) \quad \ldots \ldots \ldots \ldots \ldots \ldots \ldots (2)$$

A higher value of PSNR is good because of the superiority of the signal to that of the noise.

MSE and PSNR values of an image are between original image and stego image. The following tabulation shows the comparative study based on Wavelet alpha-blending techniques of single decomposition level.

**Table 1**

| Alpha | PSNR | |
|---|---|---|
| | Original Image Vs. Stego Image | Secret Image Vs. Extracted Secret Image |
| 0.1 | **29.0663** | 10.7082 |
| 0.2 | 22.9878 | 11.7336 |
| 0.3 | 19.4286 | 12.8906 |
| 0.4 | 16.9680 | 14.2321 |
| 0.5 | 15.0816 | 15.7956 |
| 0.6 | 13.4447 | 17.7538 |
| 0.7 | 12.1143 | 20.2454 |
| 0.8 | 10.9473 | 23.7729 |
| 0.9 | 9.9391 | **29.7718** |

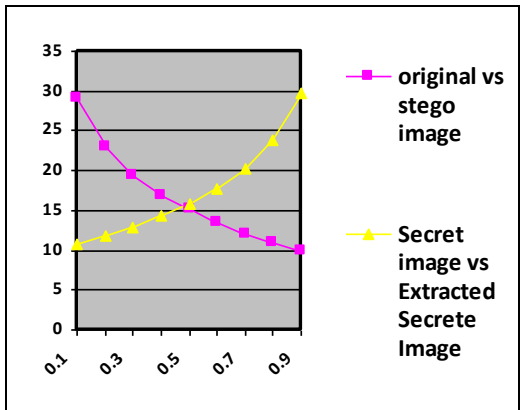

**Figure 4**

Best result for the stego image is obtained at alpha value 0.1 and best result for extracted true colour secret image is obtained at alpha value 0.9

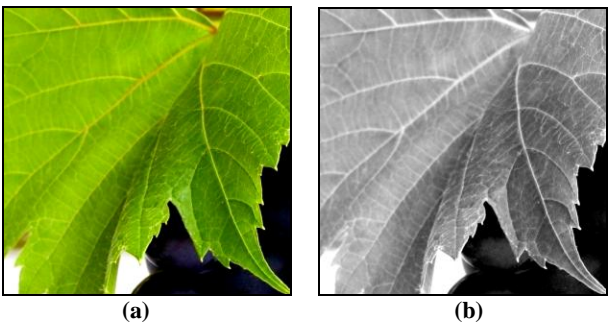

(a)    (b)

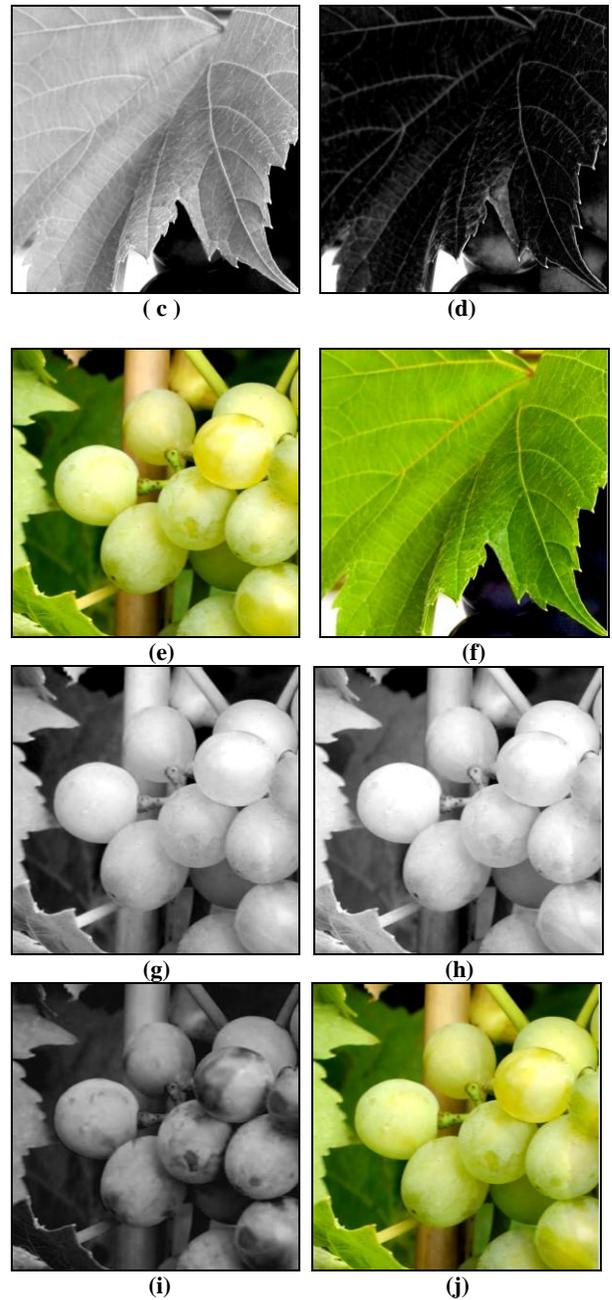

(c)    (d)

(e)    (f)

(g)    (h)

(i)    (j)

**Figure 5.  (a) Cover Image (b) R-Plane (c) G-Plane (d) B-plane (e) Secret Image (Alpha=0.9) (f) Stego Image (g) Extracted Secret Image (R-Plane) (h) Extracted Secret Image(G-Plane) (i) Extracted Secret Image (B-plane) (j) Extracted Secret RGB plane (Alpha=0.1)**

## 6. CONCLUSION

In the proposed method the color image is hidden in the different sub bands of the cover image's separate color planes. So there is a small visual change in between cover image and stego image. But due to strong security aspects this small amount of imperceptibility is acceptable. This approach can be applied for audio Steganography also, because DWT is applicable for any digital signal.